\documentclass[twocolumn,times]{aastex62}  %should match prefix of .cls file
\usepackage{graphicx}
\usepackage{float}
\usepackage{color}
\usepackage{soul}
\usepackage{xspace}

\newcommand{\Msun}{\ensuremath{\mathrm{M}_\odot}\xspace}

\begin{document}

\title{Features of Accretion-phase Gravitational-wave Emission from Two-dimensional Rotating Core-collapse Supernovae}

\author[0000-0002-4983-4589]{Michael A. Pajkos}
\affiliation{Department of Physics and Astronomy, Michigan State University, East Lansing, MI 48824, USA}
\affiliation{Department of Computational Mathematics, Science, and Engineering, Michigan State University, East Lansing, MI 48824, USA}
\affiliation{Joint Institute for Nuclear Astrophysics-Center for the Evolution of the Elements, Michigan State University, East Lansing, MI 48824, USA}

\author[0000-0002-5080-5996]{Sean M.~Couch}
\affiliation{Department of Physics and Astronomy, Michigan State University, East Lansing, MI 48824, USA}
\affiliation{Department of Computational Mathematics, Science, and Engineering, Michigan State University, East Lansing, MI 48824, USA}
\affiliation{Joint Institute for Nuclear Astrophysics-Center for the Evolution of the Elements, Michigan State University, East Lansing, MI 48824, USA}
\affiliation{National Superconducting Cyclotron Laboratory, Michigan State University, East Lansing, MI 48824, USA}

\author[0000-0002-1473-9880]{Kuo-Chuan Pan}
\affiliation{Department of Physics and Institute of Astronomy, National Tsing Hua University, Hsinchu 30013, Taiwan}

\author[0000-0002-8228-796X]{Evan P. O'Connor}
\affiliation{Department of Astronomy and The Oskar Klein Centre, Stockholm University, AlbaNova, SE-106 91 Stockholm, Sweden}

\accepted{2019 April 28 for publication in The Astrophysical Journal }

%\revised{September 27, 2016}
%\accepted{\today}
%% Command to document which AAS Journal the manuscript was submitted to.
%% Adds "Submitted to " the arguement.

%\submitjournal{ApJ}
\shorttitle{Gravitational Waves from CCSNe}
\shortauthors{Pajkos et al.}

% \abstract{}{}{}{}{} 
% 5 {} token are mandatory
 \begin{abstract}
  We explore the influence of progenitor mass and rotation on the gravitational-wave (GW) emission from core-collapse supernovae, during the postbounce, preexplosion, accretion-phase.  We present the results from 15 two-dimensional (2D) neutrino radiation-hydrodynamic simulations from initial stellar collapse to $\sim$300 ms after core bounce.  We examine the features of the GW signals for four zero-age main sequence (ZAMS) progenitor masses ranging from 12 \(M_\odot\) to 60 \(M_\odot\) and four core rotation rates from 0 to 3 rad s$^{-1}$. We find that GW strain immediately around core bounce is fairly independent of ZAMS mass and---consistent with previous findings---that it is more heavily dependent on the core angular momentum.  At later times, all nonrotating progenitors exhibit loud GW emission, which we attribute to vibrational g-modes of the protoneutron (PNS) star excited by convection in the postshock layer and the standing accretion shock instability (SASI).  We find that increasing rotation rates results in muting of the accretion-phase GW signal due to centrifugal effects that inhibit convection in the postshock region, quench the SASI, and slow the rate at which the PNS peak vibrational frequency increases.  Additionally, we verify the efficacy of our approximate general relativistic (GR) effective potential treatment of gravity by comparing our core bounce GW strains with the recent 2D GR results of other groups. 
\end{abstract}

%
%________________________________________________________________

\section{Introduction}

Core-collapse supernovae (CCSNe) became the first extra-solar multimessenger objects when SN 1987A was detected by the Kamiokande II experiment and Irvine-Michigan-Brookhaven water Cerenkov detector in 1987 \citep{bionta:1987,hirata:1987} along with concurrent electromagnetic (EM) observations \citep[see][]{arnett:1989}. With the recent detection of a neutron star merger--GW170817--in both photons and gravitational waves (GWs) by the LIGO and Virgo collaborations \citep{abbott:2016}, we have entered the era of GW multimessenger astronomy.  
So far, only the mergers of black hole binaries and a neutron star binary have been detected in GWs, but CCSNe are also predicted to be prodigious GW sources, although not quite as ``loud'' as compact object binary mergers.
Accurate predictions of the expected GW signal from CCSNe is key to increasing the likelihood of detection by GW observatories such as Advanced LIGO (aLIGO) and Advanced Virgo (AdV) and will be crucial in our ability to extract physical meaning from a future CCSN GW detection \citep{abdik:2014,gossan:2016}.

CCSNe are routinely observed in the EM window, and the data-collecting power of synoptic surveys such as the Large Synoptic Survey Telescope and Zwicky Transient Facility may increase the volume of such data for CCSNe by orders of magnitude \citep{ivezic:2019,bellm:2019}.
Still, until the late nebular phase, which is often too dim to be easily observed for distant CCSNe, the EM emission arises from the very outermost layers of the progenitor star and the central core regions, where the explosion is driven, are obscured. 
This makes it challenging to connect EM emission from CCSNe directly to the mechanism that powers them.
Due to their relatively small interaction probabilities with matter, both neutrinos and GWs offer windows through which to peer directly into the heart of a CCSN explosion.  
Moreover, these observations have broader astrophysical applications: restricting nuclear equations of state, verifying angular momentum transport in plasmas, and better understanding stellar rotation.

An observation of either GWs or neutrino emission from a nearby CCSN combined with multiband observations would allow us to place unique constraints on the physics of the explosion mechanism and key nuclear physics, such as the nuclear equation of state (EOS).  
There has yet to be a single astrophysical object detected via all three of these messengers.  
Albeit a rare event, a Galactic CCSN offers the perfect opportunity to observe such a multimessenger ``trifecta.''  
In order to increase our chances of ``hearing'' such an event in GWs, and in order to be able to extract the greatest scientific meaning from them, we need accurate predictions for CCSN GW signals from the wide range of initial conditions that give rise to these stellar explosions.

Modeling GWs from CCSNe incurs all of the challenges of simulating the CCSN mechanism itself, along with a heightened emphasis on the importance of the general relativistic (GR) treatment of gravity. 
The increased expense of including a fully dynamical spacetime evolution coupled to GR dynamics \citep[see][]{ott:2009, ott:2012} can further reduce the size of the parameter space that it is feasible to explore. 
Approximations that maintain sufficient numerical accuracy become necessary in order to reduce computational cost.  
A common approach for CCSNe, particularly in 2D, is the conformal flatness condition (CFC) approximation wherein the spatial three-metric is obtained approximately from the flat spacetime three-metric.
CFC has been shown to accurately reproduce prebounce and early postbounce signals from CCSNe to within a few percent when compared with direct solutions to Einstein's field equations  \citep{ott:2007}.  Likewise, while some differences appear, \citet{shibata:2004} find good qualitative agreement between the effective GR potential and CFC.  This conformal flatness approach has also been extended to an ``augmented CFC" scheme as introduced by \citet{saijo:2004}, refined by \citet{cordero-carrion:2009}, and utilized by \citet{bmuller:2019}.
A further approximation, also common in simulations of the CCSN mechanism, is to couple an effective GR gravitational potential to otherwise Newtonian dynamics \citep{rampp:2002, marek:2006, bruenn:2016,  moro:2018, oconnor:2018}.  
This relativistic effective potential empirically satisfies the solution to hydrostatic equilibrium according to a modified Tolman-Oppenheimer-Volkoff equation \citep{rampp:2002, marek:2006}.
This approach further reduces the computational expense of CCSN simulations relative to the CFC approach and reproduces fairly accurately gross features of CCSN simulations \citep{marek:2006, muller:2012,oconnor:2018}.  

After the infalling matter from collapse reaches nuclear densities, the core nuclei dissolve into nucleons and, eventually, the strong force becomes repulsive, halting the material infall.  On the time scale of tens of microseconds, the subsonic inner core encounters the supersonic outer core, forming a shock front.  As this shock front photodissociates overlying material and releases an enormous neutrino flux, it leaves behind a negative entropy gradient \citep{mazurek:1982,bruenn:1985,bruenn:1989}.  This scenario is unstable according to the Ledoux criterion, causing prompt convection in the postshock region \citep{burrows:1992}, therefore creating an associated emission of gravitational radiation \citep{marek:2009b,ott:2009}.  This prompt convection is an important feature that occurs in simulations that incorporate either GR or Newtonian treatments of gravity during the early postbounce phase \citep{muller:2017,richers:2017,nagakura:2018}.

Early research into GW emission from CCSNe focused on the bounce and early postbounce phase of the explosion in rotating progenitors.
These investigations found that increasing the angular momentum of the core leads to a larger strain peak at bounce \citep{muller:1982,moench:1991,yamada:1995,zwerger:1997,dimm:2002,kotake:2003,shibata:2004}.  
More recent investigations of rotating core collapse examine the role of the angular momentum distribution within the supernova progenitor and find it is only important in the rapid rotation regime, where the ratio of kinetic to gravitational potential energy ($T/|W|$) $ \gtrsim 8\%$ at bounce \citep{abdik:2014}. In order to examine GW emission at later times, different groups have considered other factors for nonrotating cases---for example, convection in the postshock region \citep{burrows:1996,muller:1997,muller:2004,marek:2009b,murphy:2009}, the standing accretion shock instability (SASI) \citep{blondin:2003,blondin:2006,ohnishi:2006,foglizzo:2007,scheck:2008,iwakami:2009,fernandez:2010}, and protoneutron star (PNS) vibrational modes \citep{cerda-duran:2013,torres-forne:2018,torres-forne:2019}.  \citet{moro:2018} investigate GW emission for moderate rotational speeds ($\Omega_{\mathrm{core}} = 0.2$ rad s$^{-1}$) for a single progenitor mass ($13 M_\odot$) over 1 second postbounce.  \citet{pan:2018}, \citet{kuroda:2018}, \citet{cerda-duran:2013}, and \citet{ott:2011} investigate the relationship between black hole formation and GW emission, for a nonrotating 40 $M_\odot$, a nonrotating  70 $M_\odot$, a rotating 35 $M_\odot$, and a rotating 75 $M_\odot$ progenitor, respectively.  These studies also find stronger GW emission at bounce with increased progenitor angular momentum and loud GW emission at later times for nonrotating CCSNe.

In this work, we present 15 axisymmetric (2D) neutrino radiation-hydrodynamic CCSN simulations.  
Our parameter space spans four progenitor masses ranging from $12\,M_\odot$ to $60\,M_\odot$ \citep{Suk:2016} and four peak core rotation speeds: $0-3 \text{ rad s}^{-1}$.  
We examine the variation in key features of the GW emission from CCSNe at these different masses and rotation rates.
Rapid rotation rates up to 2 and 3 rad s$^{-1}$ are likely rare in typical massive stars at solar metallicity due to efficient transport and loss of angular momentum \citep{heger:2005}.
While \citet{woosley:2006} observe that only 1\% of massive stars may reach the rapid rotation regime, there are high uncertainties in the stellar mass loss and magnetic braking calculations \citep{smith:2014}.  Moreover, albeit a small percentage, \citet{demink:2013} show the distinct possibility of rapidly rotating stars formed from binary interactions. Thus, there is some likelihood of rapidly rotating supernova progenitors in the mass range we explore.

In addition to the breadth of parameter space we cover, we also explore the role of rotation up to 300 ms postbounce.  
We find that rotation restricts the growth of SASI by centrifugally flattening the shock, leaving it slightly oblate. Likewise, the positive angular momentum gradient created by the rotation stabilizes the postshock convection according to the Solberg-H{\o}iland stability criterion \citep{endal:1978,fryer:2000}.  Not only are the SASI and postshock convection contributions to the gravitational radiation diminished, but the PNS vibrational signals are damped because of less turbulent downflows of matter onto the PNS surface.  
This results in a ``muting'' of the GW signal with increasing rotation speeds.
While the origins of this muting are physical, such behavior may not be seen in full 3D simulations of CCSNe due to the appearance of spiral modes of the SASI and magnetorotational instabilities (MRI) \citep{cerda-duran:2007,andresen:2019} . 

Compared with previous works, the strength of this project is its ability analyze GWs from multiple progenitors hundreds of milliseconds postbounce while accurately accounting for rotation and neutrinos.  The wide breadth of parameter space we examine allows us to reveal certain rotational effects on the GW signal in the context of a controlled study.

In the present simulations we use an approximate, effective GR potential \citep{marek:2006,oconnor:2018}.
In order to validate this approximate approach for studying GWs from CCSNe, we compare our results to those of \citet{richers:2017}, who use a CFC GR approach.  We find that our simulations produce nearly identical GW bounce signals to those of \citet{richers:2017}.  

This paper is organized as follows:  in Section \ref{sec:method} we present our methods and treatment of microphysics within our \texttt{FLASH} simulations.  We present a new method for applying initial rotation to the progenitor.  Because each progenitor evolves at a different rate and we terminate our simulations at 300 ms postbounce, we refrain from asserting which explode.  Rather, in Section \ref{sec:results}, we begin by addressing the shock front evolution for each of the progenitor masses and initial rotation velocities.  We then verify our gravitational treatment by comparing our bounce signal to GR simulations.  We explore the effect of rotation on GWs emitted hundreds of milliseconds after core bounce and discuss implications on their detectability.  In Section \ref{sec:summary} we conclude and summarize the influence of rotation on GWs from initial collapse to 300 ms postbounce. 

%__________________________________________________________________

\section{Methods and Simulation Setup}
\label{sec:method}
We utilize the \texttt{FLASH} (version 4) multiscale, multiphysics adaptive mesh refinement simulation framework for our simulations \citep{fryxell:2000,dubey:2009}.\footnote[7]{\url{http://flash.uchicago.edu/site/}}  We employ a modified, GR, effective potential \citep{marek:2006, oconnor:2018} incorporated into the multipole Poisson solver of \citet{couch:2013a}, where we retain spherical harmonic orders up through 16.   We utilize the SFHo EOS in all of our 15 simulations \citep{steiner:2013}.  Our grid setup is a 2D cylindrical geometry with the PARAMESH (v.4-dev) library for adaptive mesh refinement  \citep{macneice:2000}.  The outer boundary is $10^4$ km in all directions, with nine levels of refinement, yielding a finest grid spacing of about 0.65 km.
The maximum allowed level of refinement is decreased as a function of spherical radius, $r$, in order to maintain a resolution aspect ratio, $\Delta x_i / r$, of about 0.01, corresponding approximately to an ``angular'' resolution of $0.5^{\circ}$.

%M1 section
Neutrinos play a vital role in CCSNe.  Directly after collapse, they provide an avenue through which the PNS can cool.  As the shock propagates outward, they also provide heating in the gain region that is crucial in reviving the explosion, according to the neutrino heating mechanism.  The opacity of the material to these outflowing neutrinos must be carefully accounted for in an energy-dependent way.  We incorporate a multidimensional, multispecies, energy-dependent, two-moment scheme with an analytic closure, or the so-called M1 scheme.  Our implementation is based on \citet{oconnor:2015}, \citet{shibata:2011}, and \citet{cardall:2013}.  A detailed outline of the M1 implementation in \texttt{FLASH} is in \citet{oconnor:2018}.  This combination of rotation and the M1 neutrino treatment is similar to the work of \citet{obergaulinger:2017} and \citet{obergaulinger:2018}.  In order to reduce computational costs to explore the wide parameter space for our study, we neglect velocity-dependent neutrino transport and do not account for inelastic neutrino-electron scattering.
We use 12 energy bins spaced logarithmically up to 250 MeV, and the full set of rates and opacities we use is described in \citet{oconnor:2017a}. 
Specifically, we use the effective, many-body, corrected rates
for neutrino-nucleon, neutral current scattering of \citet{horowitz:2017}.  

We use the 12, 20, 40, and 60 \Msun nonrotating, solar-metallicity progenitors models from \citet{Suk:2016} for the present work.

\begin{figure}[t]
    \centering
    \includegraphics[scale=0.45]{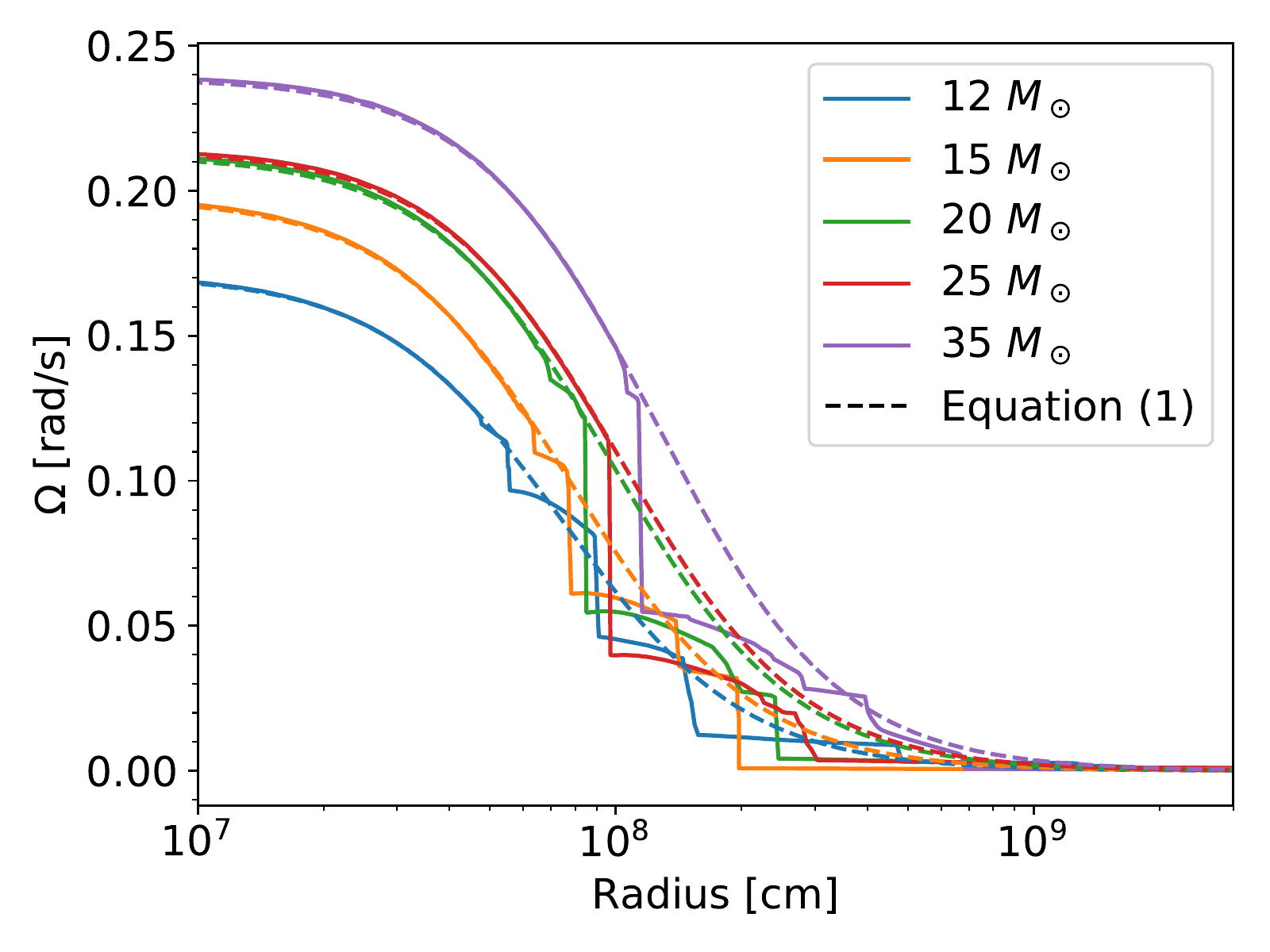}
    \caption{The rotation profiles for five of the \citet{heger:2005} progenitors.  Each solid line represents \citeauthor{heger:2005}'s (\citeyear{heger:2005}) model, and the dashed lines are Equation (\ref{eq:omega}) applied to the respective progenitor, with the appropriate differential rotation parameter.  }
    \label{fig:ovsr}
\end{figure}

\subsection{Treatment of Rotation}

The progenitor models we use are evolved without rotation.
At the start of core collapse, when we map the 1D models into our 2D grid, we apply an artificial rotation profile
\begin{equation}
    \Omega(r) = \Omega_0 \bigg[1 + \bigg(\frac{r}{A}\bigg)^2 \bigg]^{-1}, 
    \label{eq:omega}
\end{equation}

where $r = \sqrt{R^2 + z^2}$ is the spherical radius for a given cylindrical radius $R$ and altitude $z$, $\Omega_0$ is the central angular speed of the star, and $A$ is the differential rotation parameter \citep{eriguchi:1985}.  For large values of $A$, the stellar rotation is nearly solid body, whereas small values of $A$ lead to a more differential profile. 
The {\it linear} rotational velocity is then calculated by multiplying the angular speed with the distance from the rotation axis, $v_\phi (R, z) = R \Omega (r) $. 

The precise rotation rates and profiles of massive stellar cores at collapse are uncertain.
Previous work \citep[e.g.,][]{abdik:2014} treated the differential rotation parameter $A$ as a free parameter and explored the impact of its variation.
Examining the stellar evolution models of \citet{heger:2005}, which include angular momentum transport due to the Tayler-Spruit dynamo \citep{spruit:2002}, we find that $A$ is strongly determined by the {\it compactness} \citep{oconnor:2011} of the stellar core.
In order to demonstrate this, we fit the rotation profiles of the 20 progenitor models from \citet{heger:2005} to Equation (\ref{eq:omega}) in order to determine the best-fit $A$.
The models of \citet{heger:2005} include stars of zero-age main sequence (ZAMS) masses 12, 15, 20, 25, and 35 $M_{\odot}$, with various angular momentum transport parameters and initial ZAMS rotation rates.  
Using the \texttt{curve\_fit} function (in the \texttt{scipy.optimize} library) available in \texttt{Python}, we obtain $A$ values that correspond to the most accurate fits of Equation (\ref{eq:omega}) to the rotation profiles of these models.  Figure \ref{fig:ovsr} displays the radial, rotation profile for five of the aforementioned progenitors, compared with our implementation of Equation (\ref{eq:omega}), with the best-fit $A$ value.

\begin{figure}[t]
    \centering
    \includegraphics[scale=0.45]{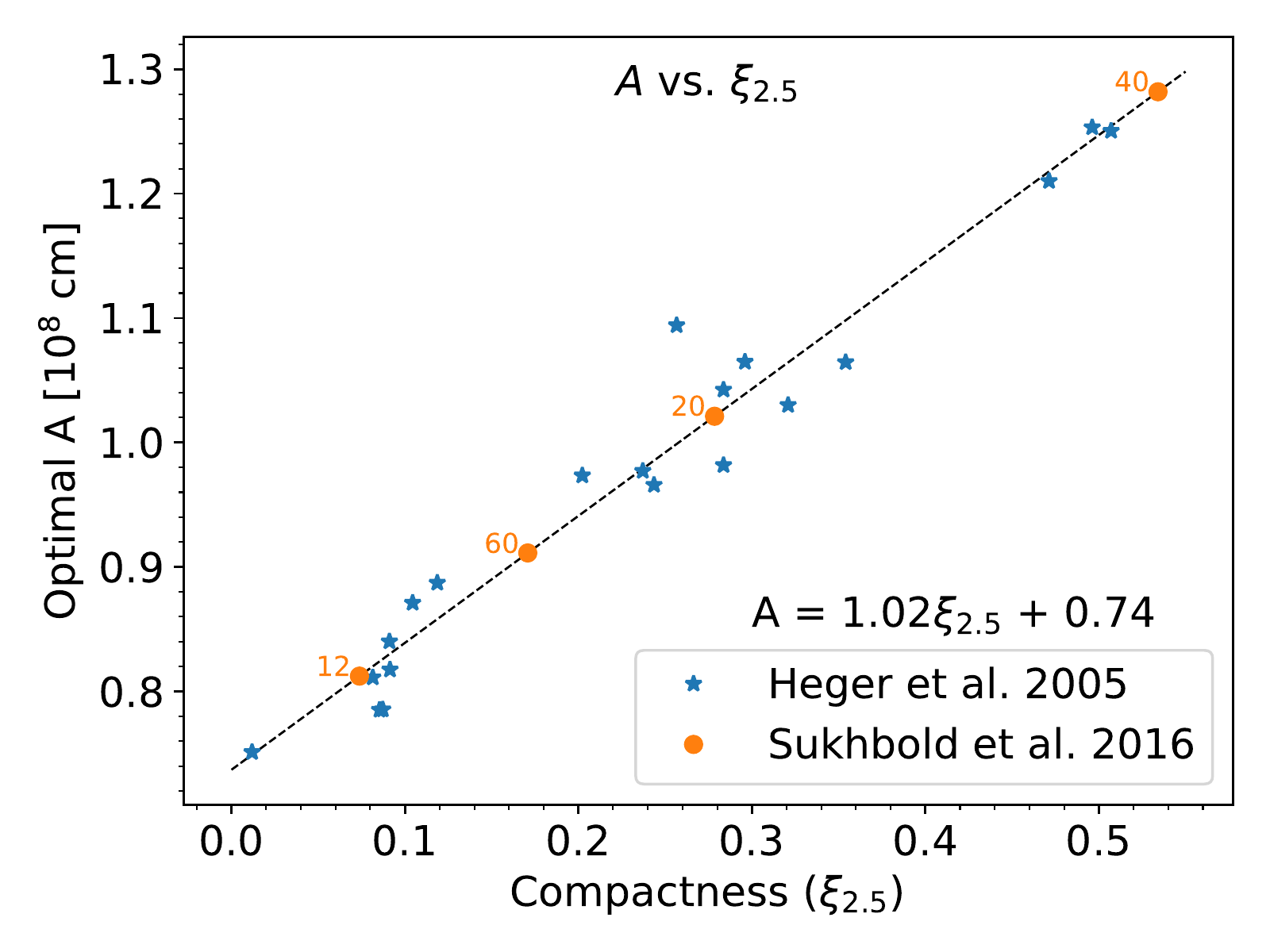}
    \caption{Linear relation between differential rotation parameter, $A$, and compactness parameter of the inner 2.5 $M_\odot$, $\xi_{2.5}$.  The linear trend is constructed from the \citet{heger:2005} rotation profiles.  We then apply the relation to the compactness values from \citet{Suk:2016} to yield the differential rotation parameters.  The progenitor ZAMS masses are labeled in units of $M_\odot$ for each respective point.}
    \label{fig:a_vs_comp}
\end{figure}

The core compactness as introduced by \citet{oconnor:2011} is defined as
\begin{equation}
    \xi_M = \left.\frac{M/M_{\odot}}{R(M_\mathrm{bary}=M)/1000 \text{km}}\right\vert_\mathrm{collapse} ,
\end{equation} 
where we choose $M = 2.5 M_\odot$, and $R(M_{\mathrm{bary}}=M) $ as the radius at which the internal baryonic mass is $2.5M_\odot$, at collapse.  
Figure \ref{fig:a_vs_comp} shows the compactness parameters from the \citet{heger:2005} models (blue stars) plotted against their best-fit $A$ values.
A clear linear relation exists between $A$ and $\xi_{2.5}$.  

\begin{table}[t]
\begin{tabular}{c|c|c}
Progenitor Mass ($M_\odot$) & Compactness & \textit{A}(1000 km) \\
\hline
12  & 0.0738 &         0.8123             \\
20  & 0.2785 &         1.021            \\
40  & 0.5341 &         1.282           \\
60  & 0.1708 &         0.9112          
\end{tabular}
\caption{Listed values for ZAMS Mass, Compactness Calculated from \citet{Suk:2016}, and Differential Rotation Parameter \textit{A}.}
\label{table:compact}
\end{table}

Using this relationship, we calculate optimal $A$ values for the four \citet{Suk:2016} progenitors we use in this work (orange circles in Figure \ref{fig:a_vs_comp}).  
The full list of progenitor masses, compactness values, and \textit{A} values we use is given in Table ~\ref{table:compact}.  

As a note, we choose to omit the 40 $M_\odot$ progenitor at $\Omega_0 = 3$ rad s$^{-1}$ from our following analysis with a numerically motivated rationale.  The $\xi_{2.5}$ value of this progenitor is nearly double that of the 20 $M_\odot$ progenitor (the next closest compactness value).  This fact displays that the 40 $M_\odot$ has a much larger differential rotation parameter compared with the other progenitors, resulting in a nearly solid-body rotation of the core. This endows the core in the 40 \Msun model with drastically more angular momentum than the other models.  The vast amount of angular momentum ultimately led to numerical instabilities in our calculations; thus, we omit the 40 $M_\odot$, with $\Omega_0 = 3$ rad s$^{-1}$ from our analysis.  \\
%So much so, in fact, core bounce is significantly delayed and occurs at much lower central densities due to the strong centrifugal barrier.
%We outline in greater detail the resulting distortion of the GW signal in Section \ref{sec:results}.  Thus, in order to preserve a maximum upper limit on the angular momentum (within a 1.75 $M_\odot$ mass radius) of $\sim 2.4\times 10^{49}$ erg s, we omit the 40 $M_\odot$, with $\Omega_0 = 3$ rad s$^{-1}$ from our analysis.  \\

%V_aniso
%Solberg Hoiland Instability

% 1) 12,15,20,25,35 fit XXX profiles to toy model
% 2) plot optimal A values to compactness from XXX
% 3) plot linear relationship
% 3) Extrapolate to higher masses (w/ given compactness) from XXX

%
%______________________________________________________________

\section{Results}
\label{sec:results}

To extract the GW signal from our simulations, we adopt the dominant, quadrupole moment formula for the gravitational strain, through the slow motion, weak-field formalism %\citep[eg.][]{misner:1973,murphy:2009}, 
\citep[eg.][]{blanchet:1990,finn:1990}
\begin{equation}
    h_+ \approx \frac{2G}{Dc^4}
    \frac{d^2I_{zz}}{dt^2},
\label{eq:quad}
\end{equation}
where $I_{zz}$ is the reduced-mass quadrupole moment, $G$ is the gravitational constant, $c$ is the speed of light, and $D$ is the distance to the source (our fiducial value is $D=10$ kpc) and we assume optimal source orientation---GWs emitted from the equator of the CCSN.\\
\par When plotting the amplitude spectral density (ASD) of the GW signal we compute the discrete Fourier transform consistent with \citet{anderson:2004} and LIGO's implementation
\begin{equation}
\widetilde{h}_{+k} = \sum^{N-1}_{j=0} h_{+j} e^{-i2\pi jk/N}
\label{eq:dft}
\end{equation}
where $i=\sqrt{-1}$.

To quantify the strength of convection within our simulations, we characterize the anisotropic velocity of the fluid motion within the postshock region according to \citet{takiwaki:2012}:

\begin{equation}
    v_{\mathrm{aniso}} = \sqrt{\left<\rho\Big((v_r - \left<v_r\right>)^2 + v_\theta^2 + v_\phi^2\Big) \right>/\left<\rho\right>}
\label{eq:vaniso}
\end{equation}
where $\left<\right>$ represents an angle average, $v_r$ is the radial velocity, $v_\theta$ is the velocity component in the polar direction, $v_\phi$ is the velocity component in the azimuthal direction, and $\rho$ is the density.  

With the introduction of rotation, a positive angular momentum gradient can be established, leading to inhibited convection, according to the Solberg-H{\o}iland stability criterion.  To quantify this criterion we calculate the condition at the equator for stability in the vertical direction, $R_{\mathrm{SH}}$, consistent with \citet{heger:2000}:

\begin{equation}
    R_{\mathrm{SH}} := \frac{g}{\rho}\Bigg[\Bigg(\frac{d\rho}{dr}\Bigg)_{\mathrm{ad}}-\frac{d\rho}{dr}\Bigg] + \frac{1}{r^3}\frac{d}{dr}(r^2\omega)^2 \geq 0
\label{eq:SHI}
\end{equation}
where $g$ is the local gravitational acceleration, $\rho$ is the density, $(d\rho/dr)_{\mathrm{ad}}$ is the radial density gradient at constant entropy and composition, $r$ is the distance from the axis of rotation, and $\omega$ is the rotational velocity. 

To examine the shape of the shock front, $R_S(\theta,\phi)$, we represent it as a linear combination of spherical harmonics, $Y_l^m(\theta,\phi)$:

\begin{equation}
    R_S(\theta,\phi) = \sum_{l=0}^\infty \sum_{m=-l}^{l} a_l^m Y_l^m(\theta,\phi)
\end{equation}
\begin{equation}
    Y_l^m = \sqrt{\frac{2l+1}{4\pi}\frac{(l-m)!}{(l+m)!}}P_l^m(\cos(\theta)) e^{im\phi}
\end{equation}
where $P_l^m$ are the associated Legendre polynomials \citep{burrows:2012,takiwaki:2012}.  However, because of the 2D nature of our simulations $\phi = 0$ and all $m = 0$ as well; thus the coefficients $a_l^0$ are

\begin{equation}
    a_l^0 = \int_0^\pi d\theta \sin(\theta)R_S(\theta)Y_l^0(\theta) .
\label{eq:sho_coeff}
\end{equation}
It follows that $a_0^0$ corresponds to the average shock radius.
\subsection{Rotation's Influence on Shock Front Evolution}

While our focus in the present work is on the GW signals up to 300 ms postbounce, we briefly discuss the impact of rotation on the evolution of the shock front as it propagates outward.  In certain cases, independent of the mechanism, the shock front may require over 300 ms to revive and complete a successful explosion.  Because our simulations are only run until 300 ms postbounce, we refrain from asserting which progenitors successfully explode.  Rather, we remark on how the average shock radii develop with time.

Of our 15 simulations, only the nonrotating 20 \Msun and 60 \Msun progenitors show substantial shock expansion.  The effect rotation has on reviving the shock is not a simple one. 
In one respect, one expects greater centrifugal support to lead to a larger shock front.  However, there are two factors that inhibit the shock from propagating outward.  The first is the inhibited convection due to the positive angular momentum gradient within the progenitor. Weaker convection results in less efficient neutrino heating \citep{dolence:2013, murphy:2013} and less positive support from turbulence in the gain region \citep{couch:2015a, mabanta:2018}.  The  second rotational element that inhibits explosions is the lack of neutrino production.  Rotation centrifugally supports matter that is infalling during the initial collapse of a star.  As such, the collapsing material does not settle as deeply into the gravitational potential of the stellar core, thereby releasing less gravitational binding energy.  This process results in a lower neutrino luminosity and slower contraction of the PNS \citep{summa:2018}. 
These two dominant effects, weaker convection and reduced neutrino luminosity, can create an unfavorable scenario for a supernova explosion that is revived by neutrino heating.

\begin{figure}[t]
    \centering
    \includegraphics[width = 0.5\textwidth]{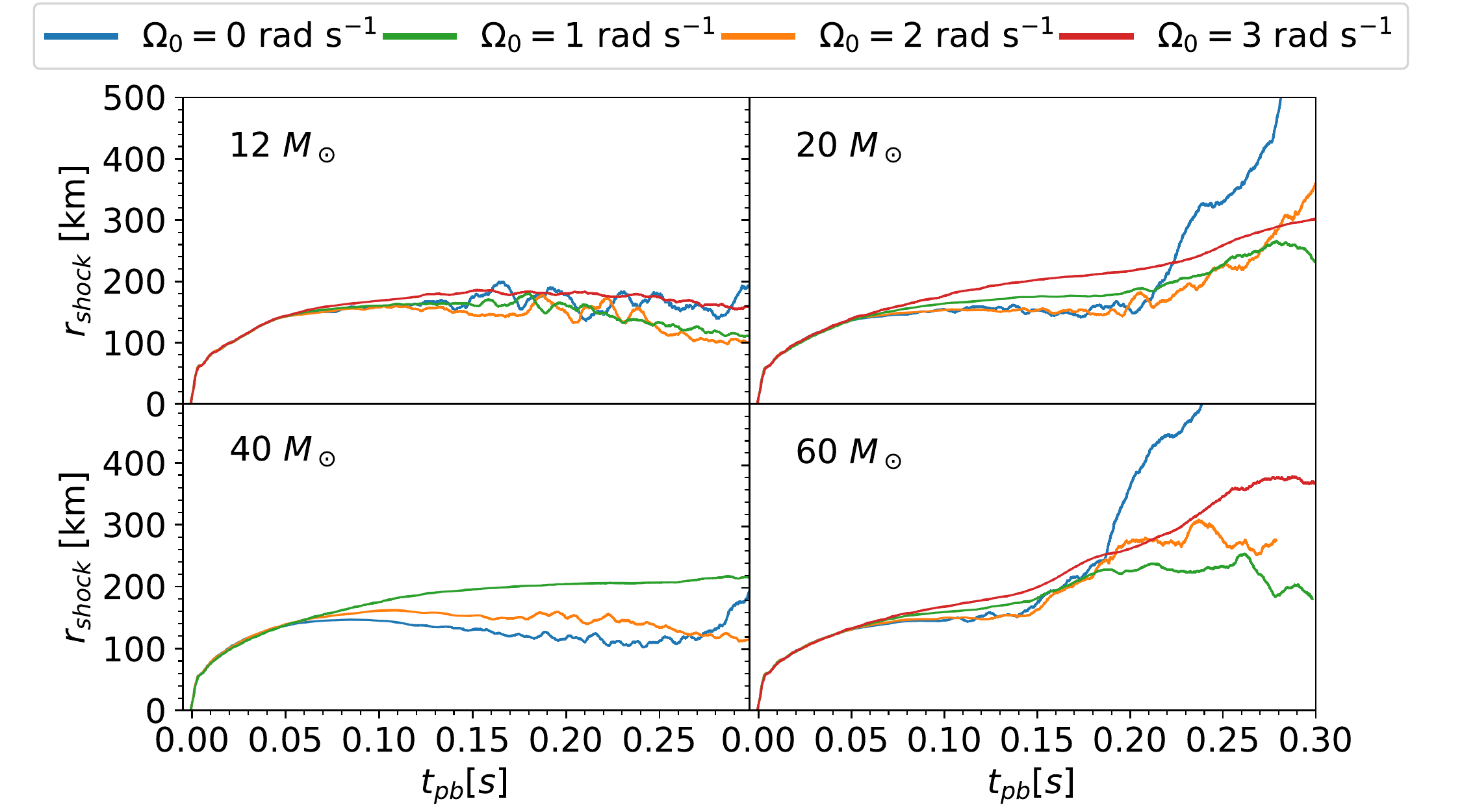}
    \caption{Shock radius evolution of the four progenitor models versus time (postbounce).  As different progenitors evolve at different rates, they may not have enough time to revive their shock front within the 300 ms interim.  As such, only the nonrotating 20 \Msun and 60 \Msun progenitors show substantial shock expansion. }
    \label{fig:shock}
\end{figure}

Despite rotation inhibiting certain aspects of a successful explosion, some of our rotating models ($\Omega_0 = 3$ rad s$^{-1}$, 20 \Msun, and 60 \Msun) have advancing shock radii.  With longer simulation times, these could lead to explosion.  In these cases, it seems that rotation could be sufficiently rapid to overcome the deleterious effects on convection and reduced neutrino luminosity.
Similar nonmonotonic behavior is reported by \citet{summa:2018} in their 2D simulations.
Hence, the introduction of rotation involves competing forces that can enhance or diminish the shock.  Figure \ref{fig:shock} shows the average shock radius evolution versus time (postbounce) over our entire parameter space.

\subsection{Comparison with CFC GR}

 \begin{figure}[t]
    \centering
    \includegraphics[width=0.48\textwidth]{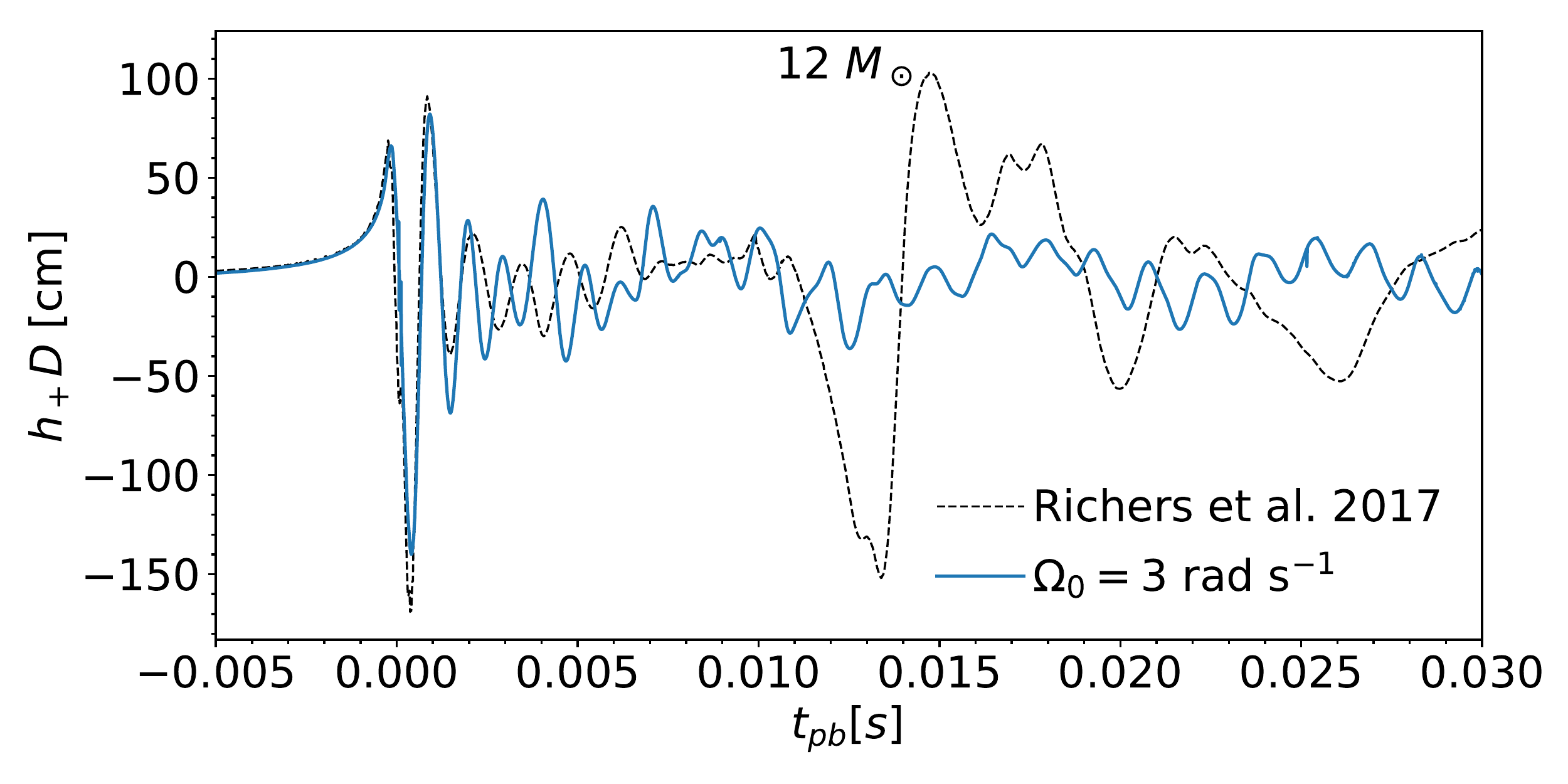}
    \caption{GW strain vs. time (postbounce) for a 12 \(M_\odot\) progenitor \citep{woosley:2007} with $\Omega_0 = 3$ rad s$^{-1}$.  Plotted in the dashed line is the GW strain from \citet{richers:2017} using the CFC \texttt{CoCoNuT} code, and the solid line is our result using the effective relativistic potential coupled with Newtonian dynamics.  While the different grids and treatment of hydrodynamics lead to differences in the strain in the early postbounce phase, we qualitatively verify our gravitational treatment by obtaining a nearly exact bounce signal. }
    \label{fig:bounce_cfc}
\end{figure}

\begin{figure*}[t]
  \centering     %%% not \center
  \includegraphics[width=0.49\textwidth]{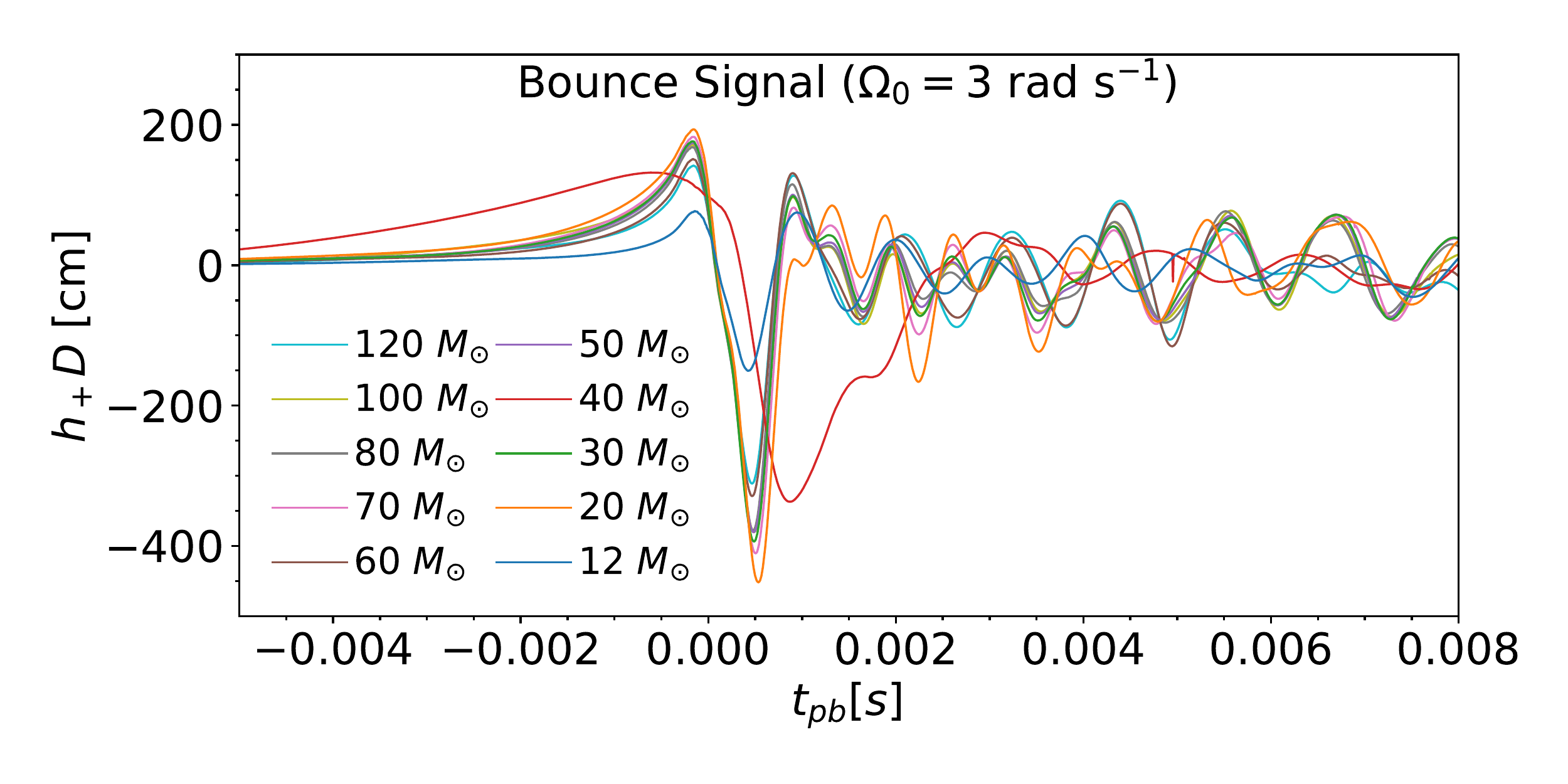}
  \includegraphics[width=0.49\textwidth]{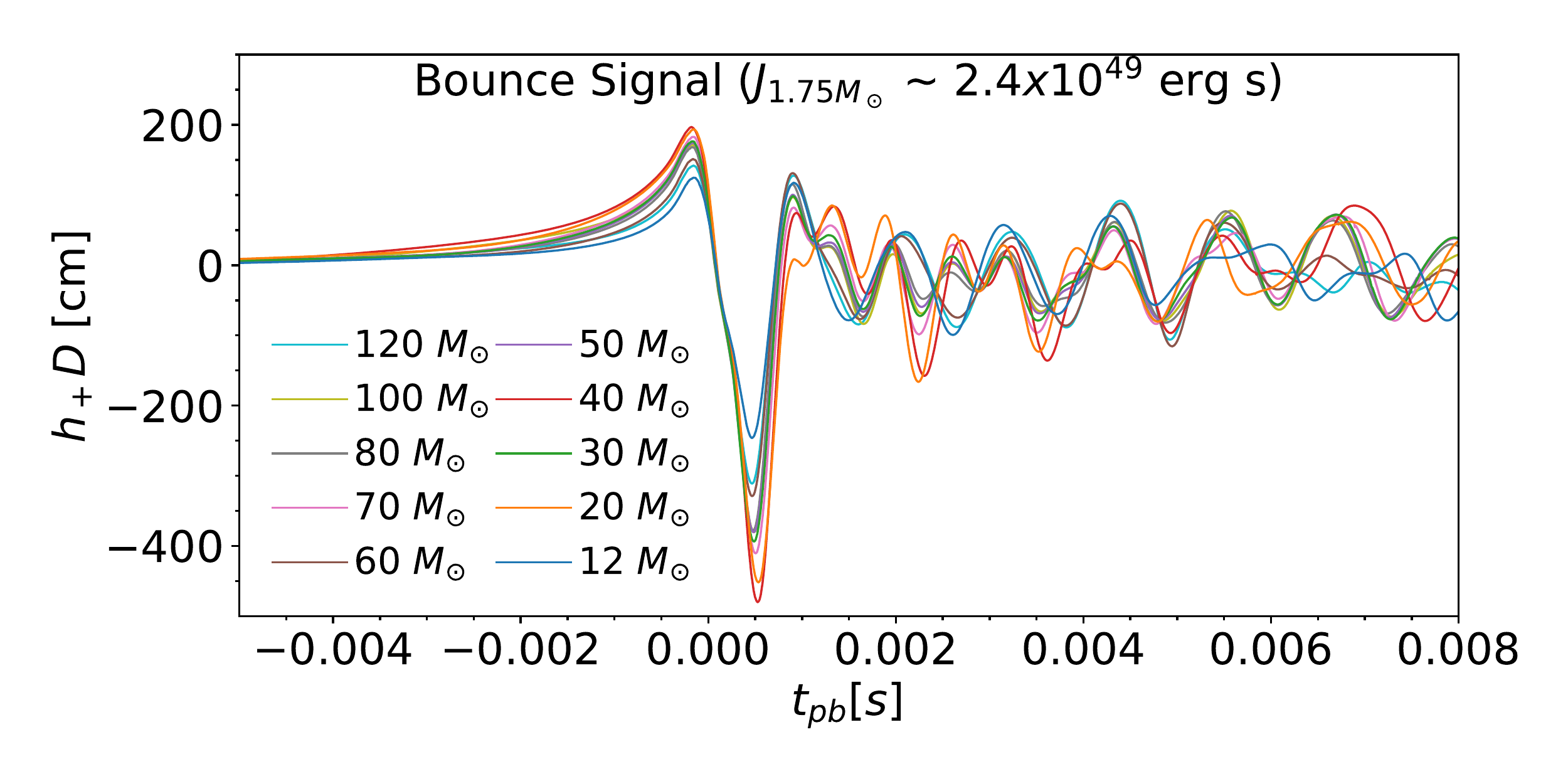}
  \caption{(Left) GW bounce signal from all 10 progenitor masses with $\Omega_0 = 3 \text{ rad s}^{-1}$.  By applying Equation (\ref{eq:omega}), we assign a radially dependent, angular velocity to our progenitors.  Because the central density profiles of each progenitor are different---namely, a less compact $12\,M_\odot$ and more compact $40\,M_\odot$---the progenitor cores are endowed with different amounts of angular momenta.   (Right) Modified bounce signals after adjusting rotation rates to yield similar angular momenta ($\sim 2.4\times10^{49} \text{erg s}$) of the inner $1.75\,M_\odot$ of matter.  As predicted by \citet{dimm:2008} and \citet{abdik:2010,abdik:2014}, the GW bounce signals depend on the inner core angular momentum at bounce, not the original ZAMS mass.}
  \label{fig:bounce}
\end{figure*}

In multidimensional simulations of CCSNe, the treatment of gravity must offer a balance between numerical accuracy and computational cost.   The CFC offers a nearly identical GW signal, compared with full GR, while reducing simulation time \citep{ott:2007}.  Figure \ref{fig:bounce_cfc} offers a qualitative check of our effective GR potential compared with CFC \citep{richers:2017}.  We incorporate an identical deleptonization profile \citep{lieb:2005} and SFHo EOS \citep{steiner:2013} for a 12 \Msun progenitor \citep{woosley:2007}.  Moreover, we match the differential rotation parameter and rotation profile by selecting an $A = 634$ km and $\Omega_0 = 3$ rad s$^{-1}$.  For this comparison, we match the neutrino physics of \citet{richers:2017}'s simulation by using a ray-by-ray, three-species, neutrino leakage scheme \citep{oconnor:2010,couch:2014}.  We capture a nearly identical bounce signal and similar strain up to 5 ms postbounce. \par
However, after the initial bounce signal ring-down, it is clear that the different computational treatments of hydrodynamics and grid geometry result in differences in the GW strains.  Although not exact, the efficiency of the effective GR potential offers a reasonable method to accurately model the GW signal from CCSNe to within 10\% and allows for larger sweeps of parameter space \citep{muller:2013}.

%-----------------------------------
%-----------------------

\subsection{ZAMS Influence on Gravitational Bounce Signal}

While different progenitors $\gtrsim$$8\, M_\odot$ will experience widely varied evolution, once their iron cores reach the effective Chandrasekhar mass \citep{baron:1990} and collapse commences, the physics of the collapse becomes somewhat universal.
In particular, the mass of the homologously collapsing inner core is fixed more by microphysics than by the macrophysics of varied stellar evolution. 
This nearly identical inner core mass across the ZAMS parameter space yields similar core angular momenta, for identical rotation rates.  Hence, the core bounce signal is nearly indistinguishable between progenitor masses.  For further verification of our gravitational treatment, we perform 12 additional simulations using neutrino leakage---from collapse---until 8 ms after core bounce, in order to replicate this bounce signal degeneracy, using the \citet{Suk:2016} progenitors.  Outlined by \citet{ott:2012}, neutrino leakage has a small effect on the GW bounce and early postbounce signal.  
Moreover, our results are consistent with 3D, fully GR predictions given by \citet {ott:2012} that similar core angular momenta yield similar GW bounce signals.  

Figure \ref{fig:bounce} displays the bounce signals for all 10 progenitor masses, ranging from 12 \Msun to 120 \Msun.  The left panel is for uniform rotational velocity prescriptions at $\Omega_0 = 3\text{ rad s}^{-1}$.  As previously highlighted, the angular momentum of the inner core is the main contributor to the gravitational bounce signal.  While many of the waveforms have similar amplitudes, there are two clear outliers: the $12\,M_\odot$ and $40\,M_\odot$ progenitors.  The $12\,M_\odot$ and $40\,M_\odot$ progenitors, respectively, have lower and higher compactness values at collapse, by nearly a factor of 2.  Because we endow each progenitor with angular velocity, and not specific angular momentum, the more compact $40\,M_\odot$ progenitor will receive more angular momentum, compared with the remaining progenitors, thereby affecting the resulting GW bounce signal.  As outlined by \citet{dimm:2008}, once a star is sufficiently rotating, the centrifugal support slows the bounce, diminishing the GW bounce amplitude and widening out the bounce peak of the waveform.  

The inverse is true for the $12\,M_\odot$ case.  Because it has a less compact inner core at collapse, using Equation (\ref{eq:omega}) leads to less initial angular momentum, thereby producing a lower amplitude bounce signal.  After modifying the initial rotation rates of both progenitors, to match the progenitor core angular momenta (right panel of Figure \ref{fig:bounce}), the change produces nearly identical GW bounce signals.  

Hence, our results from exploring the bounce signal over a wide range of progenitor masses support the results of previous studies of the angular momentum dependence of the GW signal \citep{dimm:2008,abdik:2010,abdik:2014} but also serve as a cautionary note for future groups who perform rotating CCSN simulations with a wide variety of progenitor models.  
It is worth noting that other rotational treatments exist beyond the simple angular velocity law, such as specifying a radial, specific angular momentum profile \citep[eg.,][]{oconnor:2011} or using the rotational profile from the rotating stellar evolution models directly \citep{summa:2018}.  The profiles used by \citet{oconnor:2011} lead to a roughly uniform rotation rate within a mass coordinate of 1 $M_\odot$ and $\Omega(r)$ decreasing with  $r^{2}$ outside this mass coordinate.  \citet{summa:2018} utilize two different rotation schemes: one that matches the \citet{heger:2005} models seen in Figure \ref{fig:ovsr} and one that is solid body out to $\sim 1500$ km and then falls as $r^{-3/2}$.

\subsection{Rotational Influence on Accretion-phase GW Emission}

Our results in the previous section support the efficacy of our effective GR potential for accurately modeling the GW signals from CCSNe.  While the effective GR potential has been shown to overestimate peak frequency from GWs compared with GR, it produces similar GW amplitudes and accurately captures PNS compactness during the accretion-phase \citep{muller:2013}.  Thus we now turn to exploring the rotational effects on the GW signal during the accretion-phase, up to 300 ms after bounce.

While the consistency of the inner core mass for a collapsing iron core creates a setting where envelope mass has little effect on the bounce signal, the postbounce dynamics of the explosion largely depend on the mass surrounding the PNS.  For nonrotating CCSNe, the shock front propagates outward and loses energy due to dissociation of iron nuclei and neutrino cooling.  In the case of rotation, the initial progenitor and resulting shock front become more oblate.  Rotation can affect the GW emission in three respects: the postshock convection is damped, the SASI becomes restricted, and it slows the rate at which the PNS peak vibrational frequency increases.

\begin{figure}[]
    \centering
    \includegraphics[width=0.5\textwidth]{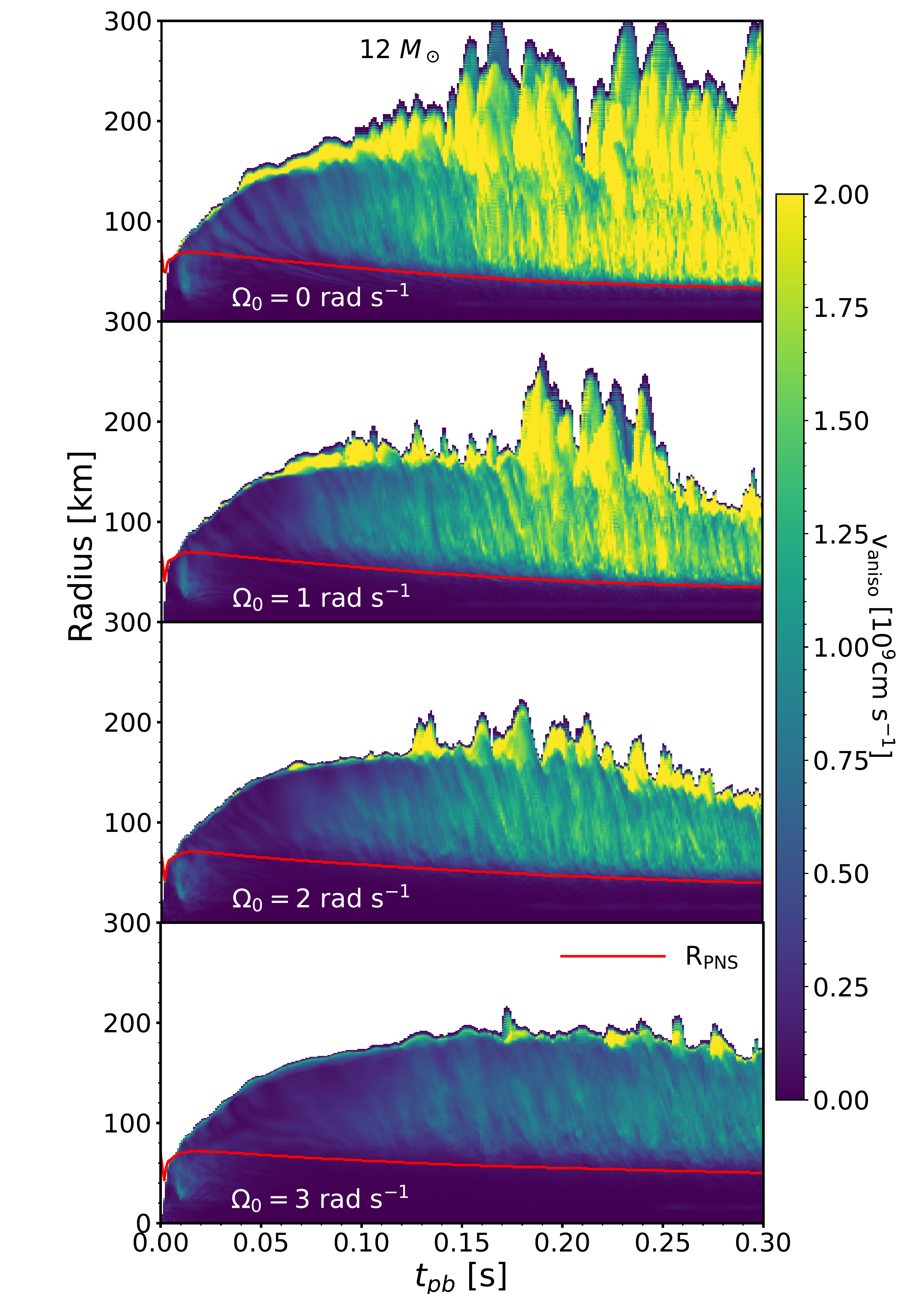}
    \caption{Spherically averaged anisotropic velocity of the postshock region for the 12 \Msun progenitor.  Brighter colors correspond to increased convection in the postshock region according to Equation (\ref{eq:vaniso}).  As rotational velocity increases, convective activity is inhibited.  Traced in red is the radius of the PNS.}
    \label{fig:vaniso}
\end{figure}

\begin{figure}[]
    \centering
    \includegraphics[width=0.5\textwidth]{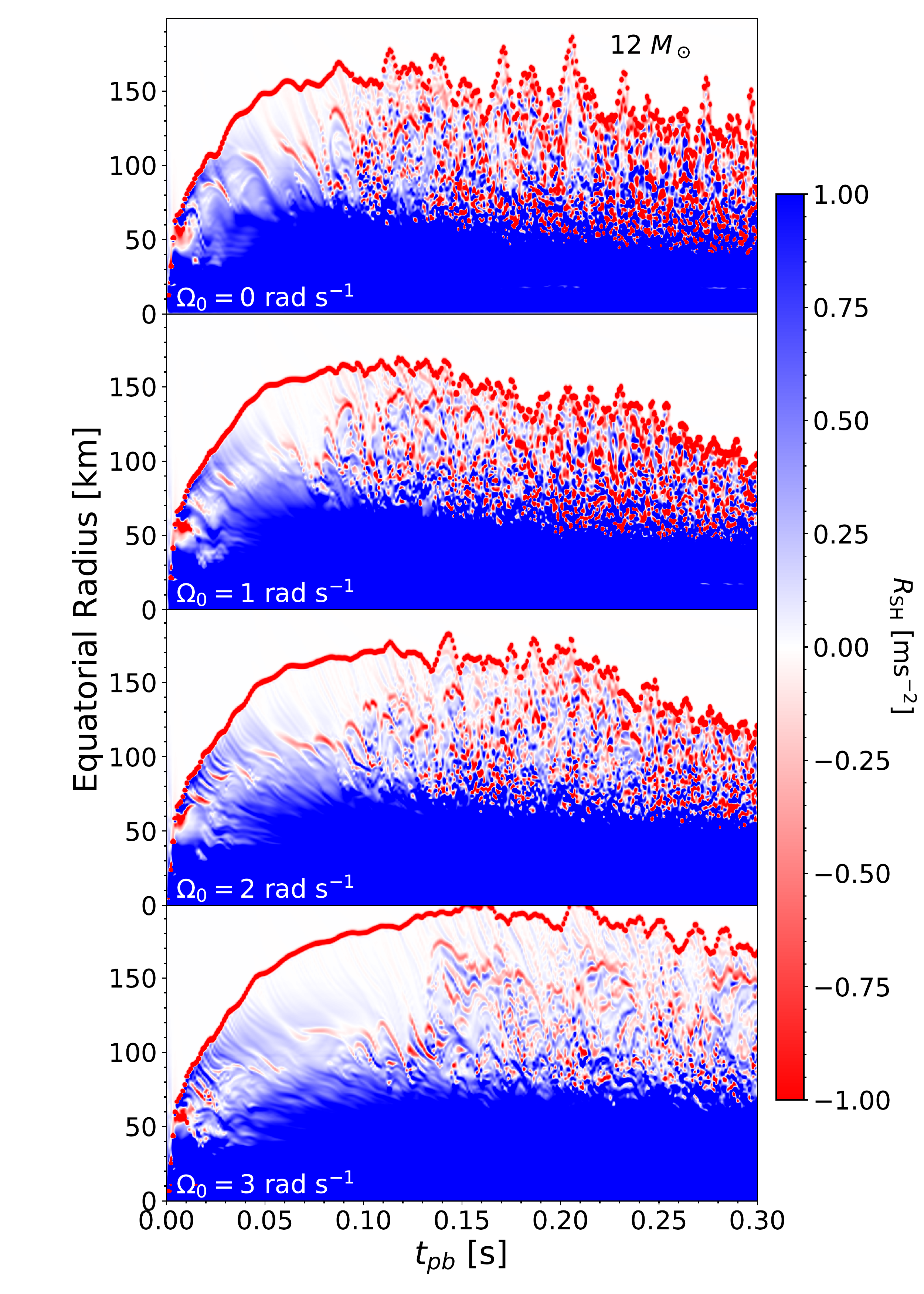}
    \caption{Slices along the equator of the 12 \Msun progenitor at each rotational velocity.  Colors correspond to the Solberg-H{\o}iland stability criterion, $R_{\mathrm{SH}}$, from Equation (\ref{eq:SHI}).  As rotational velocity increases, not only does the convectively stable band in the core grow (seen in blue), but the amount of convection within the postshock region (seen in red) decreases as well.  The differences in shock radius evolution between Figure \ref{fig:vaniso} and this figure arise because Figure \ref{fig:vaniso} uses an angular average over the domain, whereas this figure uses equatorial slices.}
    \label{fig:SHI}
\end{figure}

 \begin{figure*}[htp]
  \centering     %%% not \center
  \includegraphics[width=\textwidth]{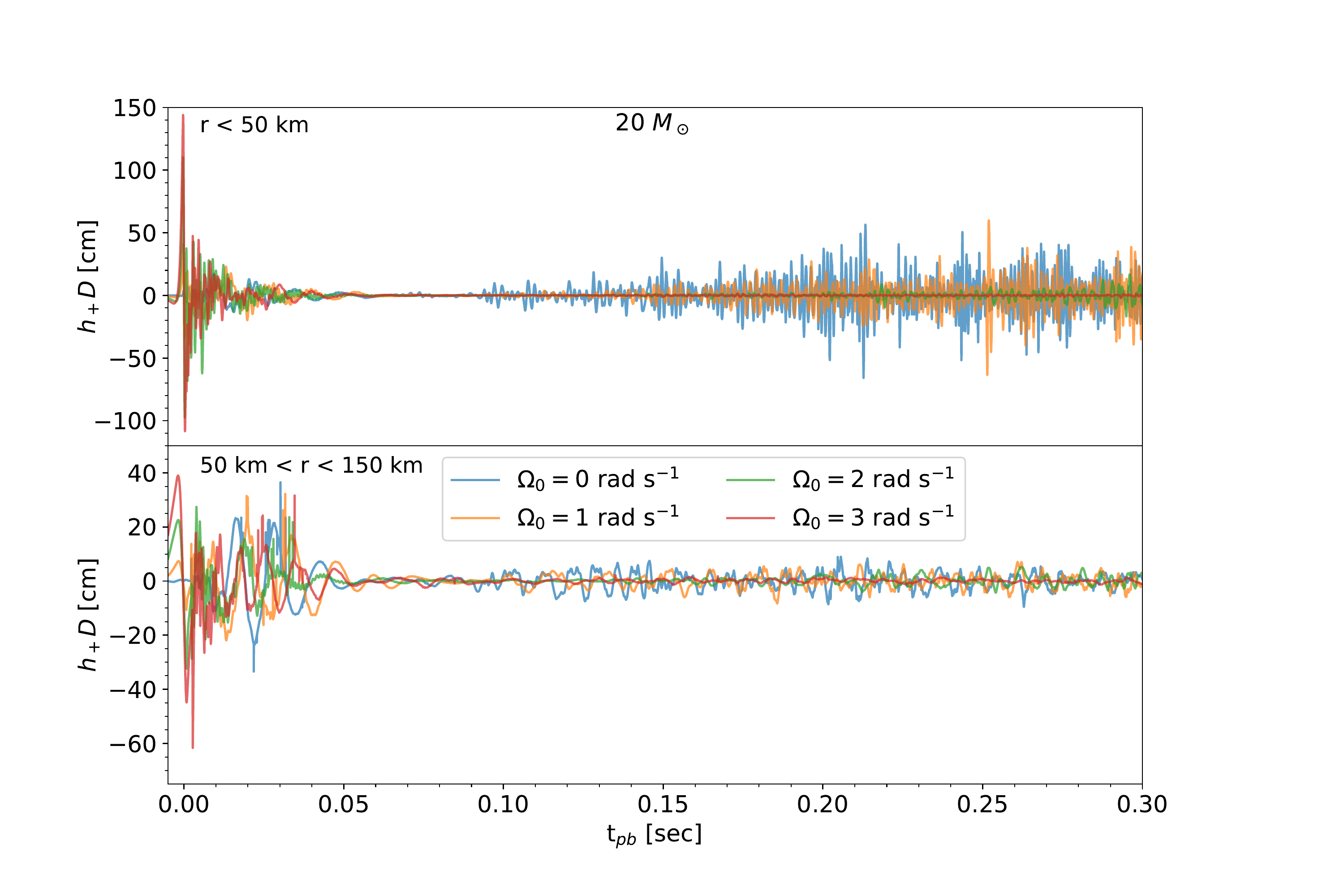}
  \caption{Time domain waveforms for the 20 $M_\odot$ progenitor.  Each panel corresponds to the region from which the GWs are emitted.  The large contribution in the top panel indicates the main source of GWs during the accretion-phase is from the vibrating PNS.  The lower panel displays the inhibited convective signal $\sim 50$--$100 $ ms postbounce that is characteristic of this quiescent phase.}
  \label{fig:region}
\end{figure*}

\begin{figure}
    \centering
    \includegraphics[width=0.5\textwidth]{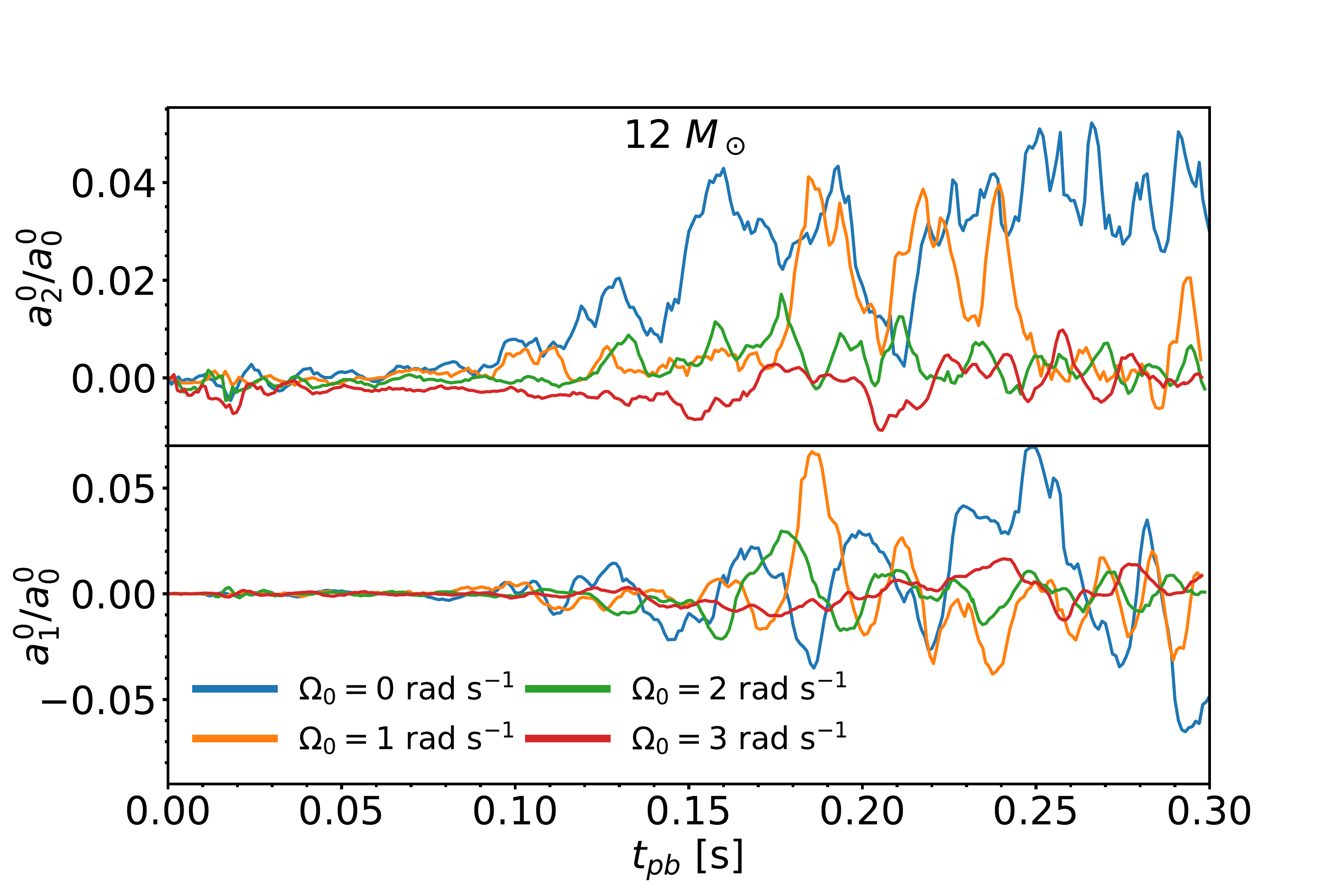}
    \caption{Coefficients from spherical harmonic decomposition of the shock front, outlined in Equation (\ref{eq:sho_coeff}).  The $a_1^0/a_0^0$ and $a_2^0/a_0^0$ terms describe the overall dipole and quadrupole nature of the shock front, respectively.  As the SASI is one of the main contributors to the creation of asymmetries in the shock front, the lower $a$ values correspond to a less prolate shock, or one with diminished SASI.}
    \label{fig:sasi}
\end{figure}

  As the $\Omega_0$ value increases in our models, a positive angular momentum gradient is established within the postshock region, partially stabilizing it to convection via the Solberg-H{\o}iland instability criterion \citep{endal:1978,fryer:2000}.  We quantify the reduced convection in Figure \ref{fig:vaniso}.  Brighter colors correspond to higher values of the anisotropic velocity as outlined in Equation (\ref{eq:vaniso}).  As expected, the convection in the gain region is reduced with increasing rotational velocity.  To tie this inhibited convection to the Solberg-H{\o}iland instability criterion, we follow the prescription of Section 2.3.2 of \citet{heger:2000}.  We quantify this instability criterion as outlined in Equation (\ref{eq:SHI}) by taking slices along the equator and tracking its evolution.  Figure \ref{fig:SHI} displays the $R_{\mathrm{SH}}$ value along the equator of the 12 \Msun progenitor for all four rotational velocities.  As the $\Omega_0$ increases, the propensity for convection (colored red) within the postshock region clearly decreases.  This inhibited convection results in weakened turbulent mass motion within the gain region, thereby reducing the GW amplitude at later times.
 
Furthermore, we recast our analysis by focusing on regions within the CCSN that emit GWs.  The lower panel of Figure \ref{fig:region} displays the inhibited convective signal  with increasing rotation, as the GW signal in the gain region becomes increasingly muted.  The typical convective signals in the early postbounce regime are then quickly washed out by the postbounce ring-down of the PNS, as rotation increases.

%4 panel GW strain plot
\begin{figure*}[t]
\includegraphics[width=\textwidth]{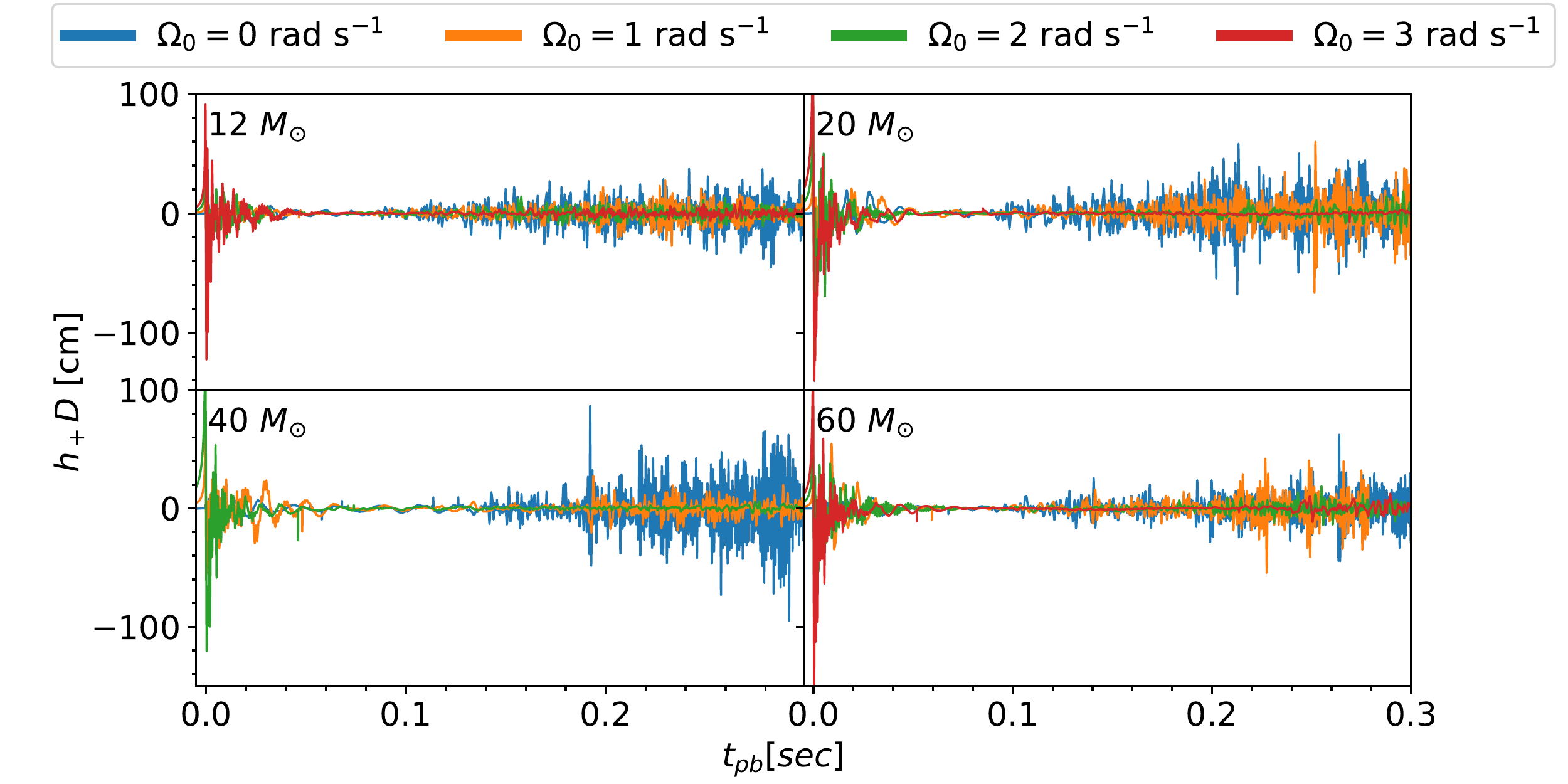}
\centering
\caption{Time domain waveforms over our entire parameter space.  For all four progenitor masses, the rotational muting of the accretion-phase GW signal is clear.  While there is some weak dependence in the character of the accretion-phase GW signals with progenitor ZAMS mass, the rotational muting occurs for all progenitors.}
\label{fig:ccsn_all}
\end{figure*}

\begin{figure}[t]
    \centering
    \includegraphics[trim=80 0 0 0, scale=0.38]{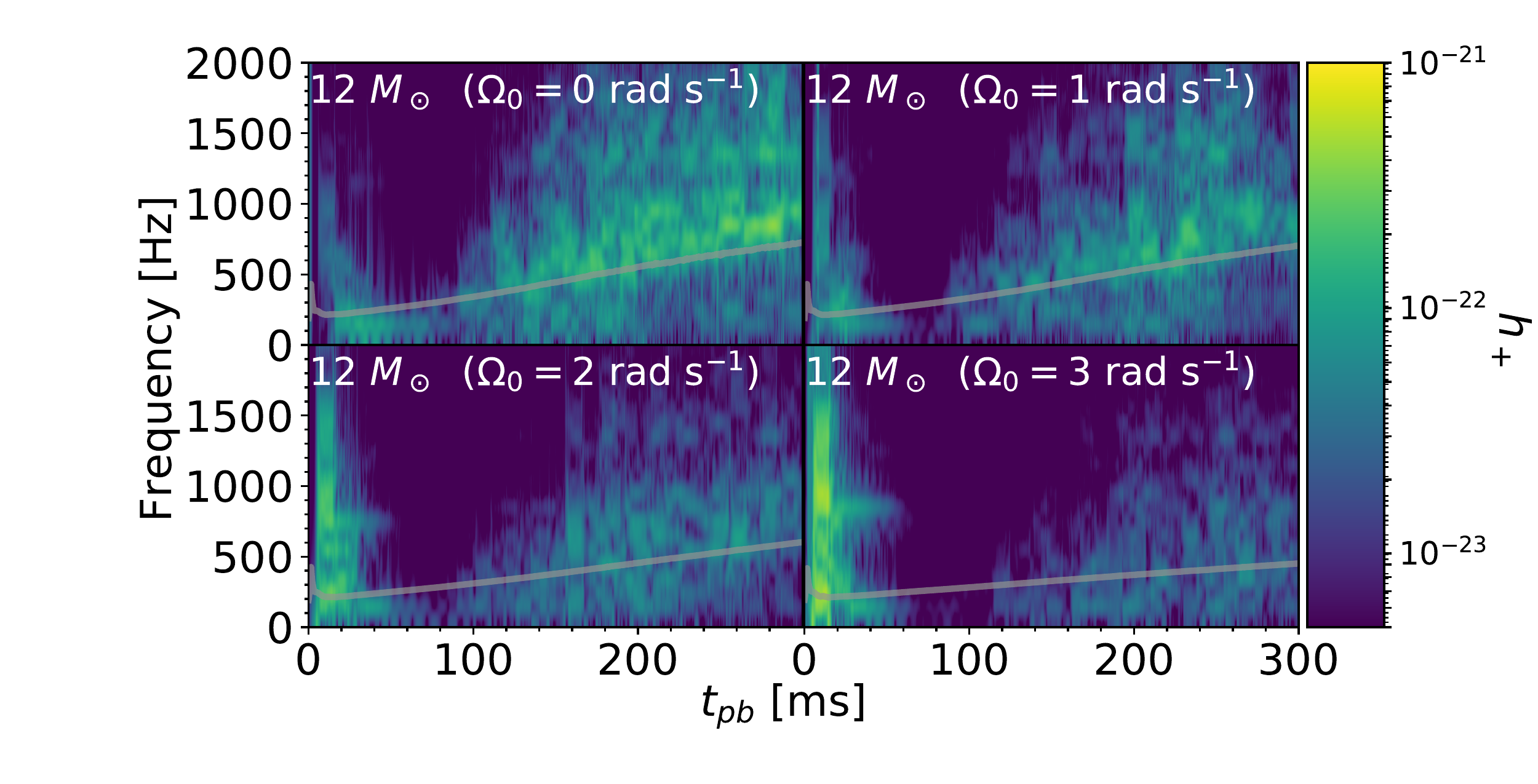}
    \caption{Spectrograms for the $12 M_\odot$ progenitor over all four rotational velocities.  The key aspects revealed by the spectrogram are the rotational muting of GWs and the flattening of the signal from the surface g-mode of the PNS.  This flattening is a product of the enlarged radius of the PNS due to centrifugal effects and can be characterized by the dynamical frequency ($f_{dyn} = \sqrt{G \overline{\rho}}$), overlaid in gray.}
    \label{fig:2x2}
\end{figure}

\begin{figure*}[t!]
  \centering     %%% not \center
  \includegraphics[width=\textwidth]{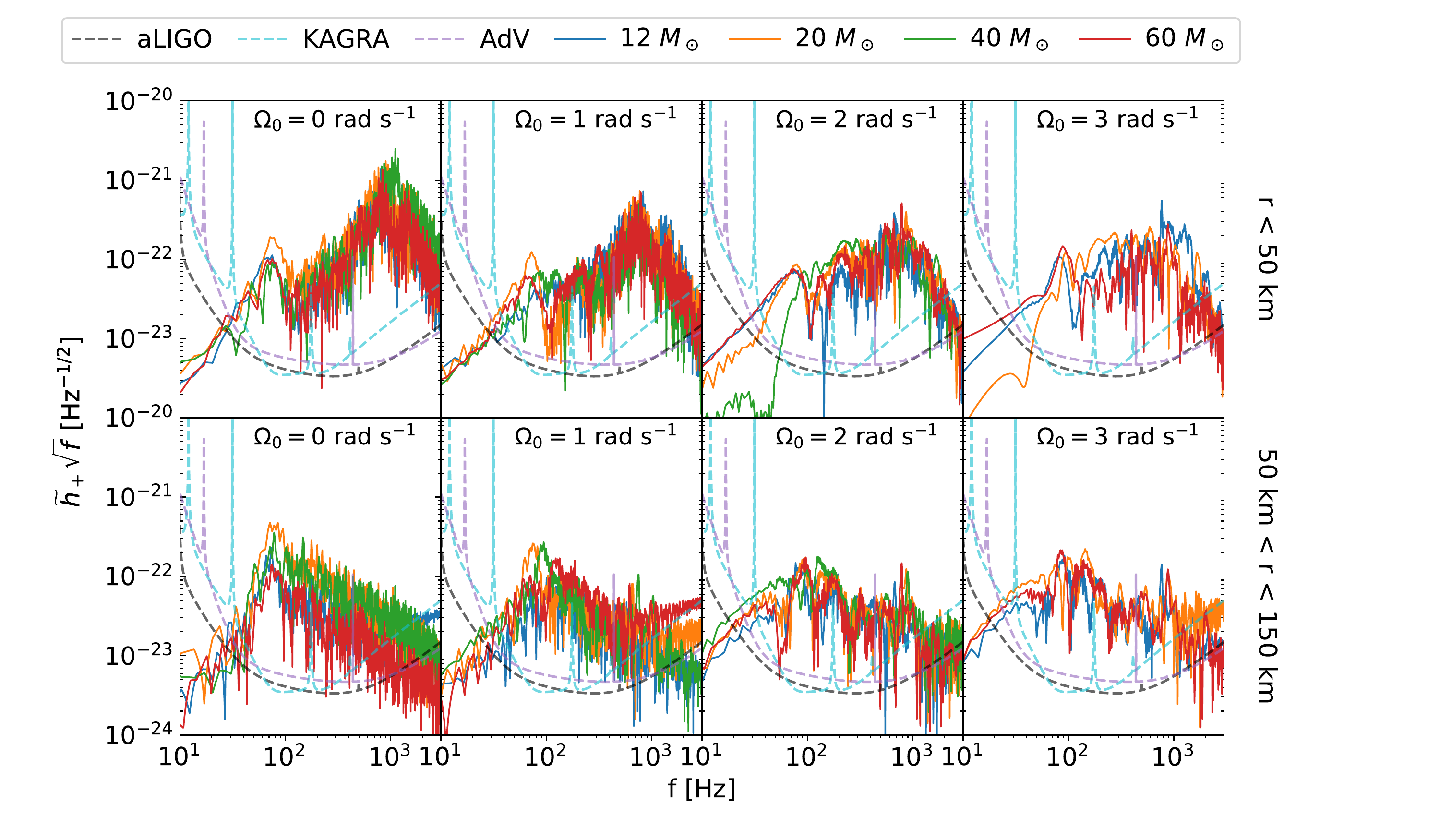}
  \caption{ASD plot of all progenitors for all rotation rates from $t_{be}+6$ ms $\rightarrow t_{be}+300$ ms, with an assumed distance of 10 kpc.  The rotational muting of the fundamental PNS g-mode is displayed as the peak frequency ($\sim 800$ Hz) becomes less prevalent, with increasing rotation rate.  Likewise, the low-frequency signals ($\sim40$ Hz) from the gain region become more audible, with increasing rotational velocity.  The damping of the vibrational modes of the PNS allows the slower postshock convection to contribute more to the overall GW signal.  Plotted in the black dashed line is the design sensitivity curve for aLIGO in the zero-detuning, high-sensitivity configuration \citep{barsotti:2018}.  The cyan dashed line is the predicted KAGRA detuned, sensitivity curve \citep{komori:2017}.  The purple dashed line is the design sensitivity curve for AdV \citep{abbott:2018}.}
  \label{fig:spetra_long}
\end{figure*}

Under nonrotating conditions, the shock can grow unstable due to nonradial deformations exciting a vortical-acoustic cycle that leads to the growth of large-scale shock asymmetries, that is, the SASI \citep{blondin:2003, blondin:2006, scheck:2008,marek:2009a}.
In 2D simulations, the SASI excites large, oscillatory flows along both poles that drive changes in entropy capable of causing postshock convection.  It is worth noting in 3D simulations that the SASI can excite `spiral' modes that correspond to nonzero $m$ values \citep{blondin:2007,kuroda:2016}.  The high degree of nonlinearity among the hydrodynamic flows, neutrino interactions, and gravitational effects can yield matter flow that is quadrupolar, thereby resulting in GW emission.  However, when the shock becomes restricted in the polar direction, due to centrifugal effects, SASI development is inhibited.  To quantify the role of SASI, we decompose the shock front into coefficients based on the spherical harmonics, $Y_l^m$, according to Equation (\ref{eq:sho_coeff}). Figure \ref{fig:sasi} illustrates the evolution of the $a_1^0$ and $a_2^0$ coefficients over time.  Both coefficients quantify the deviation of the shock from spherical symmetry.  Specifically, the $a_1^0$ term describes the overall dipole nature of the shock, and the $a_2^0$ term describes its quadrupole nature.  Both coefficients are normalized by the mean shock radius, $a^0_0$.  Clearly, both approach zero with increasing rotational velocity.  Physically, this effect corresponds to a shock that is becoming less prolate.  To further illustrate this transition, we direct the reader to Figures \ref{fig:vaniso} and \ref{fig:SHI}.  Figure \ref{fig:vaniso} takes an angular average to calculate $v_{\mathrm{aniso}}$, whereas Figure \ref{fig:SHI}, by contrast, uses equatorial slices  to calculate the Solberg-H{\o}iland stability criterion.  The boundary between the white and colored region in both panels then acts as a proxy for average shock radius and equatorial shock radius, respectively.  Thus, as rotational velocity increases, average shock radius decreases, while increasing the equatorial shock radius.  Put more simply, the rotation in our 2D simulations acts to create less prolate shock fronts.  Hence, because SASI plays a significant role in creating a shock that is extended along the axis of rotation, we conclude that the effect of SASI is reduced as rotational velocity increases in our 2D simulations.
While we expect the SASI activity to contribute uniquely to the GW spectrum, depending on progenitor mass, the rotational muting of the GWs is universal across ZAMS mass parameter space, as illustrated in Figure \ref{fig:ccsn_all}. 
Both \citet{burrows:2007}  and \citet{moro:2018} point out the partial suppression of SASI, but the former does not focus on the gravitational radiation emitted and the latter only examines a single, slow rotating, progenitor.  Our work provides strong support for the rotational muting of accretion-phase GWs, over such a wide region of parameter space of 2D CCSN simulations. 
 
With respect to PNSs, a variety of oscillatory modes exist that could be of interest to current and future GW astronomers: fundamental f-modes, pressure based p-modes, and gravity g-modes---due to chemical composition and temperature gradients \citep{unno:1989}.  The typical frequency of the PNS f-mode is around 1 kHz, and p-modes have frequencies greater than f-modes, which are of little use to GW astronomers, with the current detector capabilities \citep{ho:2018}.  
The frequencies of g-modes, however, are on the order of hundreds of hertz, falling squarely within the detectability range of current GW detectors \citep{martynov:2016}.  
The top panel of Figure \ref{fig:region} displays the contribution of the vibrating PNS to the majority of the GW signal during the accretion-phase, with $h_+D$ normalized strain amplitudes around 50 cm.  
These g-modes are thought to be excited by downflows from postshock convection or internal PNS convection \citep{marek:2009b,murphy:2009,muller:2013}.  
Figure \ref{fig:2x2} shows a spectrogram for the $12 \, M_\odot$ progenitor over all rotational speeds, where lighter colors represent greater strain amplitudes, $h_+$.  The dominant yellow band that extends from 100 to 1000 Hz represents this contribution.  
Overlaid in gray is the dynamical frequency that is characterized by the average density of the PNS, $\overline{\rho}$, and gravitational constant, $G$, $f_\mathrm{dyn} = \sqrt{G \overline{\rho}}$,  that evolves synchronously with the g-mode contribution.  The synchronized evolution of $f_\mathrm{dyn}$ and the frequency at which the PNS emits gravitational radiation are no coincidence.  As both are fundamentally related to the mass and radius of the PNS, we expect that both are affected similarly when introducing rotation.  The initial progenitor rotation will centrifugally support the PNS, thereby leaving it with a larger average radius.  Similar to two tuning forks of different lengths, the PNS with a larger radius will emit at a lower frequency, compared with a smaller PNS.  This ``flattening'' of the emitted frequency is displayed in Figure \ref{fig:2x2}.  Furthermore, Figure \ref{fig:2x2} provides a different lens through which the rotational muting is displayed, via the progressively darker panels with increasing rotational velocity.  We note that more robust peak GW frequency calculations exist \citep[e.g.,][]{muller:2013,moro:2018}, but we find that the simple $f_\mathrm{dyn}$ relation gives a good estimate of the PNS peak frequency.

We also Fourier transform the accretion-phase GW signal, as displayed in Figure \ref{fig:spetra_long} and scale the magnitude of the Fourier coefficients by $\sqrt{f}$ in order to produce ASD plots.  These plots commonly display the sensitivity curves of current and next-generation GW detectors.  We define $t_{\mathrm{be}}$ similar to \citet{richers:2017} as the third zero crossing of the gravitational strain.  We focus on the signal later than $t_{\mathrm{be}} + 6$ ms in order to remove the bounce signal and early postbounce oscillation contribution to the signal.  
The dominant contributions are the prompt convection, SASI, and surface g-modes of the PNS---as displayed by a peak frequency ranging from 700 to 1000 Hz.  Universally, the prevalence of the peak frequency decreases with increasing rotational velocity.  It is worth noting this peak could shift to higher frequencies with longer simulation times.

When incorporating magnetic fields into CCSN simulations, other instabilities may arise that can compromise stability in the postshock region and possibly affect the behavior of the PNS.  The $\alpha$--$\Omega$ dynamo and MRI are two such mechanisms that can reexcite postshock convection; however, work from \citet{bonanno:2005} suggests that the $\alpha$--$\Omega$ dynamo is unimportant on dynamical timescales.   MRI has the potential to drive convection in the postshock region, yet as the strength and geometry of magnetic fields in 3D simulations are largely still unknown, we exclude them from our simulations \citep{cerda-duran:2007}.

\subsection{Observability of the Accretion-phase Signal}

Overlaid on our ASD plots is the expected sensitivity of future GW observatories.  
In the black, cyan, and purple dashed lines we have plotted the sensitivity curves of design sensitivity for aLIGO in the zero-detuning, high-sensitivity configuration, the predicted KAGRA detuned sensitivity curve, and design sensitivity for AdV, respectively \citep{komori:2017,abbott:2018,barsotti:2018}.
These curves represent the incoherent sum of the principal noise sources to the best understanding of the respective collaborations.  While these curves do not guarantee the performance of the detectors, they act as good guides for their anticipated sensitivities nonetheless. 

Beyond the decreased prevalence of the peak frequency, an interesting trend emerges in Figure \ref{fig:spetra_long} as rotation increases.  
We separate the GW signals by region within the star.  The top row of Figure \ref{fig:spetra_long} corresponds to GWs originating from the inner 50 km of the supernova, and the GW signal in the bottom row originates from radial distances between 50 and 150 km from the supernova center.  In the top row, we note the first peak of emission, around 80 Hz, is independent of rotation.  We point to the bright, higher $v_\mathrm{aniso}$ region in Figure \ref{fig:vaniso} within the first 25 ms postbounce that is present for all rotational velocities.
Focusing on the bottom row, we highlight a noticeable difference in the amplitude of the low-frequency contributions, particularly around 40 Hz.  The nonrotating progenitors have undetectable low-frequency signals for all three detectors, whereas rotating progenitors create measurable signals at low frequencies. 
This enhanced low-frequency signal may provide an observable feature that can help determine progenitor angular momentum information.  

The amplitude of low-frequency GWs in the 50--150 km region of the supernova increases with rotational velocity, but this trend does not occur within the inner 50 km.  As such, we restrict the low-frequency GW contribution to the gain region.  We note the two main physical mechanisms in this region correspond to postshock convection and the SASI.  While both mechanisms are reduced in strength due to rotational effects, they do not completely cease.  This fact is displayed in Figure \ref{fig:vaniso}, as the region between 50 and 150 km is nonzero.  For the nonrotating case, the high convective velocities (bright yellow) create higher frequency GWs within the 100 km region of interest.  As rotation velocity increases, convective velocities decrease enough to cease exciting the vibrational modes of the PNS.  These slower convective flows thereby reduce the total amount of power produced by the GWs and push the peak GW frequency---from the gain region---to lower frequencies.  Performing an order-of-magnitude estimate on the source of the low-frequency signal, from Figure \ref{fig:vaniso}, we find  $v_{\mathrm{aniso}} \sim 1 \times 10^9$ cm s$^{-1}$ for $\Omega_0 = 0$ rad s$^{-1}$ and $v_{\mathrm{aniso}} \sim 5 \times 10^8$ cm s$^{-1}$ for $\Omega_0 = 3$ rad s$^{-1}$.  As the region of interest is $\sim 10^7$ cm, we yield an estimated frequency of emission $f_{\mathrm{low}}$ around $\sim 100$ Hz and $\sim 50$ Hz, respectively.  These quantitative frequency estimates are reflected in the ASD as the contribution from peak frequency ($\sim 100$ Hz) from the gain region decreases, while the contribution $\sim 40$ Hz increases.

% Because we plan to incorporate more detailed microphysics into our simulations, we refrain from making quantitative relations between low frequency spectra and progenitor angular momentum, in aims to conduct a more thorough investigation in future work.

\section{Summary and Conclusion}
\label{sec:summary}

The strength of this project is its ability analyze GWs hundreds of milliseconds postbounce from multiple progenitors while accurately accounting for rotation and neutrinos.  The wide breadth of parameter space we examine allows us to reveal certain rotational effects on the GW signal in the context of a controlled study.  We have explored the influence of rotation on the GW emission from CCSNe for four different progenitors and four different core rotational speeds.  
We point out that there exists a roughly linear relation between compactness, $\xi$, and the differential rotation parameter, $A$, as defined in Equation (\ref{eq:omega}). 
Using this relation, we calculate appropriate $A$ values for each progenitor mass, based on their individual compactness parameters of the \citet{Suk:2016} progenitors.  Of our 15 simulations, only two nonrotating progenitors have average shock radii that show substantial shock expansion, while the remaining rotating progenitors do not because of rotationally inhibited convection in the gain region and less neutrino production.  In agreement with other recent work \citep[e.g.,][]{summa:2018}, we find a complex interplay between centrifugal support and neutrino heating as successful explosions do not display a monotonic relationship with rotation.

While there are more accurate treatments of gravity, we utilize the effective GR potential in order to streamline calculations, granting us the ability to explore larger sections of parameter space. 
We find that our results utilizing this approximation match very closely the CCSN bounce signal of CFC gravity with GR hydrodynamics \citep{richers:2017}.  

The main contributors to the GW signal (10--300 ms postbounce) are postbounce convection, the SASI, and the surface g-modes of the PNS \citep{moro:2018}.  By establishing a positive angular momentum gradient, the convection is suppressed according to the Solberg-H{\o}iland stability criterion \citep{endal:1978,fryer:2000}.  The more oblate shock front inhibits the bipolar sloshing of the SASI.  Since the SASI and convection are the principal drivers exciting the g-modes of the PNS, vibrational emission from the PNS is also inhibited by rotation.  
We, therefore, find that rotation in 2D CCSN simulations results in the muting of GW emission.
This result is consistent across progenitors with different ZAMS masses. 

Before the PNS g-mode signal is completely muted, as rotation gradually increases, this signal is pushed to lower peak frequencies and can be characterized by its dynamical frequency.  This observation is no coincidence as both  fundamentally depend on the radius and mass of the PNS.  With more centrifugal support, the PNS has a larger radius.  This larger radius causes the surface of the PNS to emit at lower frequencies, thereby producing a ``flatter,'' lower frequency signal.

We reveal a novel rotational effect on the GW signal during the accretion-phase.  We notice that the nonrotating progenitors all produce low-frequency signals ($\sim 40$ Hz) that are below the plausible detection threshold of the aLIGO and KAGRA detectors, whereas the progenitors with larger angular velocities produce measurable GW signals in this frequency range.  We attribute this increase of low-frequency emission to the SASI and postshock convection.  For nonrotating progenitors, the convective velocity within the postshock region is high, emitting GWs $\sim 100$ Hz.  As rotational velocity increases, the PNS GW contribution is reduced.  Likewise, as the convection slows, the mass within the gain region emits at lower GW frequencies.  The slower convective flows reduce the total amount of GW power and push the peak GW frequency from the gain region to lower values.  Whereas previous rotating core-collapse GW studies have focused on the bounce signal as a means to determine rotational features, or have focused on late time signals without rotation, our study unifies both facets and opens the door to measuring GW signals beyond the bounce phase that encode progenitor, angular momentum information. 
We postpone asserting quantitative relations between low-frequency emission and progenitor angular momentum until we incorporate more detailed microphysics.

While our approximations have allowed us to make large sweeps of parameter space, they leave room for us to include more robust microphysics.  In an ideal situation, we would compute 3D simulations, including full GR, magnetohydrodynamics, and GR Boltzmann neutrino transport that incorporates velocity dependence and inelastic scattering on electrons and nucleons.  These additions would allow for more accurate gravitational waveforms and allow other phenomena to occur, for example the $m\ne 0$ (spiral) modes of the SASI. \citet{andresen:2019} recently highlighted the rotational effects on GWs in 3D.  Inherent to its 3D nature, their study finds the strongest GW amplitudes at high rotation velocities due to these spiral modes.  The 2D geometry of our study, however, allows us to observe the relative strength of the convective signal, without interference from $m\ne 0$ modes, as we extend beyond the case of a single rotational velocity.
While the physical origin of this muting that damps the convection and the SASI is not constrained only to 2D, in 3D, as \citet{andresen:2019} point out, other nonaxisymmetric instabilities can contribute to significant GW emission at late times, negating this rotational muting effect.
Thus, once again, we are reminded of the key role of 3D simulations in the study of the CCSN mechanism.

\acknowledgements

We would like to thank Jess McIver for pointing us to the aLIGO and AdV sensitivity curves.  M.A.P. was supported by a Michigan State University Distinguished Fellowship. 
S.M.C. is supported by the U.S. Department of Energy, Office of Science, Office of Nuclear Physics,
under award Nos. DE-SC0015904 and DE-SC0017955 and the \textit{Chandra
X-ray Observatory} under grant No. TM7-18005X.
This research was supported by the Exascale Computing Project (17-SC-20-SC), a collaborative effort of two U.S. Department of Energy organizations (Office of Science and the National Nuclear Security Administration) that are responsible for the planning and preparation of a capable exascale ecosystem, including software, applications, hardware, advanced system engineering, and early testbed platforms, in support of the nation's exascale computing imperative.
The software used in this
work was in part developed by the DOE NNSA-ASC OASCR Flash Center at
the University of Chicago.  

    \software{FLASH (see footnote 7) \citep{fryxell:2000,fryxell:2010}, Matplotlib\footnote[8]{\url{https://matplotlib.org/}} \citep{hunter:2007},
    NuLib\footnote[9]{\url{http://www.nulib.org}}
    \citep{oconnor:2015},
    NumPy\footnote[10]{\url{http://www.numpy.org/}} \citep{vanderwalt:2011}, SciPy\footnote[11]{\url{https://www.scipy.org/}} \citep{jones:2001}}

%\bibliographystyle{aasjournal}%name of .bst file

%\bibliography{masterDB, ref} %name of .bib file
\bibliography{ms}

%-------------------------------------------------------------------

% \begin{thebibliography}{}

\end{document}